\documentclass[11pt,twoside]{article}
\usepackage{asp2006}
\usepackage{psfig}
\usepackage{epsf}
\usepackage{graphics}
\usepackage{lscape}
\def\edcomment#1{\iffalse\marginpar{\raggedright\sl#1\/}\else\relax\fi}
\reversemarginpar

\begin{document}
\title{Micro-Sigmoids as Progenitors of Polar Coronal Jets}
\author{N.-E. Raouafi, P. N. Bernasconi and D. M. Rust}
\affil{JHU-APL, Laurel, MD, USA}
\author{M. K. Georgoulis}
\affil{RCAAM, Academy of Athens, Athens GR-11527, Greece}

\begin{abstract}
Observations from the Hinode X-ray telescope (XRT) are used to study the structure of X-ray bright points (XBPs), sources of coronal jets. Several jet events are found to erupt from S-shaped bright points, suggesting that coronal micro-sigmoids are progenitors of the jets. The observations may help to explain numerous characteristics of coronal jets, such as helical
structures and shapes. They also suggest that solar activity may be self-similar within a wide range of scales in terms of both properties and evolution of the observed coronal structures. 
\end{abstract}

\vspace{-0.5cm}
\section{Introduction}

The X-ray telescope (XRT; Golub et al. 2007) on Hinode provides data with unprecedented spatial and temporal resolution. This allows to resolve small-scale solar structures (e.g., XBPs)  and to follow their evolution preceding the eruption of coronal jets. Coronal jets often occur in conjunction with other phenomena that are likely due to magnetic reconnection, such as H$\alpha$ surges (Rust 1968; Canfield et al. 1996) and polar plumes (Raouafi et al. 2008 and Raouafi 2009).

The distinctive collimated structure of coronal jets inspired the anemone model, in which a simple dipolar magnetic structure is embedded in a background of open fields. The anemone-shaped structures are widely believed to form through the reconnection of emerging dipolar, magnetic loop systems with the open coronal field (Shibata et al. 1994). 

We took advantage of the exceptional data quality from XRT to study the morphology and fine structure of XBPs leading to coronal jets. We particularly address the relationship between polar coronal micro-sigmoids and X-ray jets.

\section{Observations and Data Analysis}

XRT images of the polar coronal holes are our primary data source to study the origins, formation, and evolution of X-ray jets. The data are corrected from instrumental effects utilizing XRT calibration procedures available on SolarSoft.

\begin{figure}[!ht]
\begin{center}
\plotone{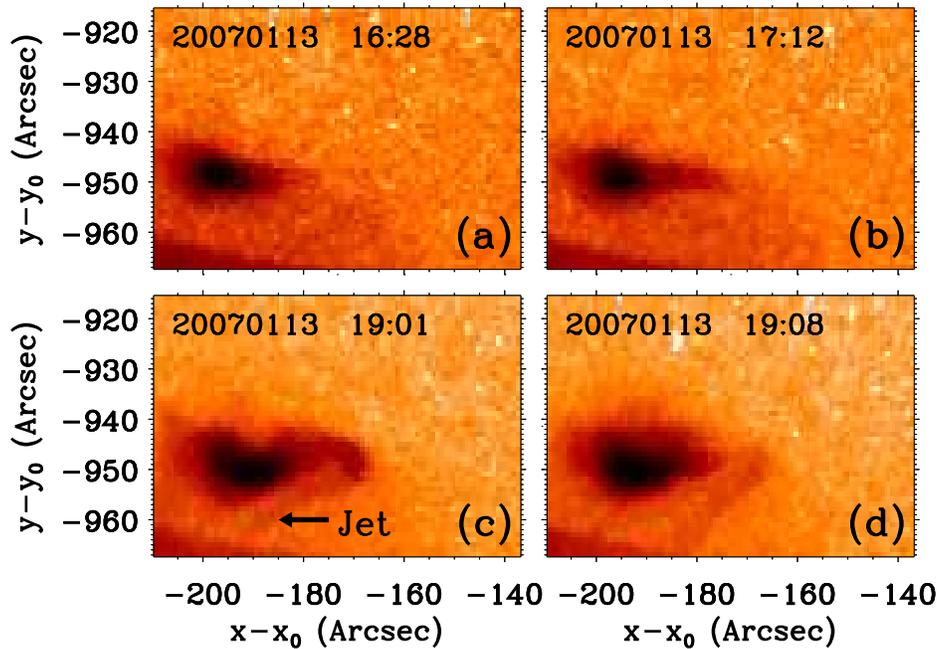}
\caption{Negative images displaying the evolution of an XBP in the southern polar coronal hole into a micro-sigmoid that yielded a jet (shown by the arrow in panel c). Notice the significant shape change of the structure in the aftermath of the jet. The observations are recorded with the XRT Al\_poly filter. \label{h3fig1}}
\end{center}
\end{figure}

\begin{figure}[!ht]
\begin{center}
\plotone{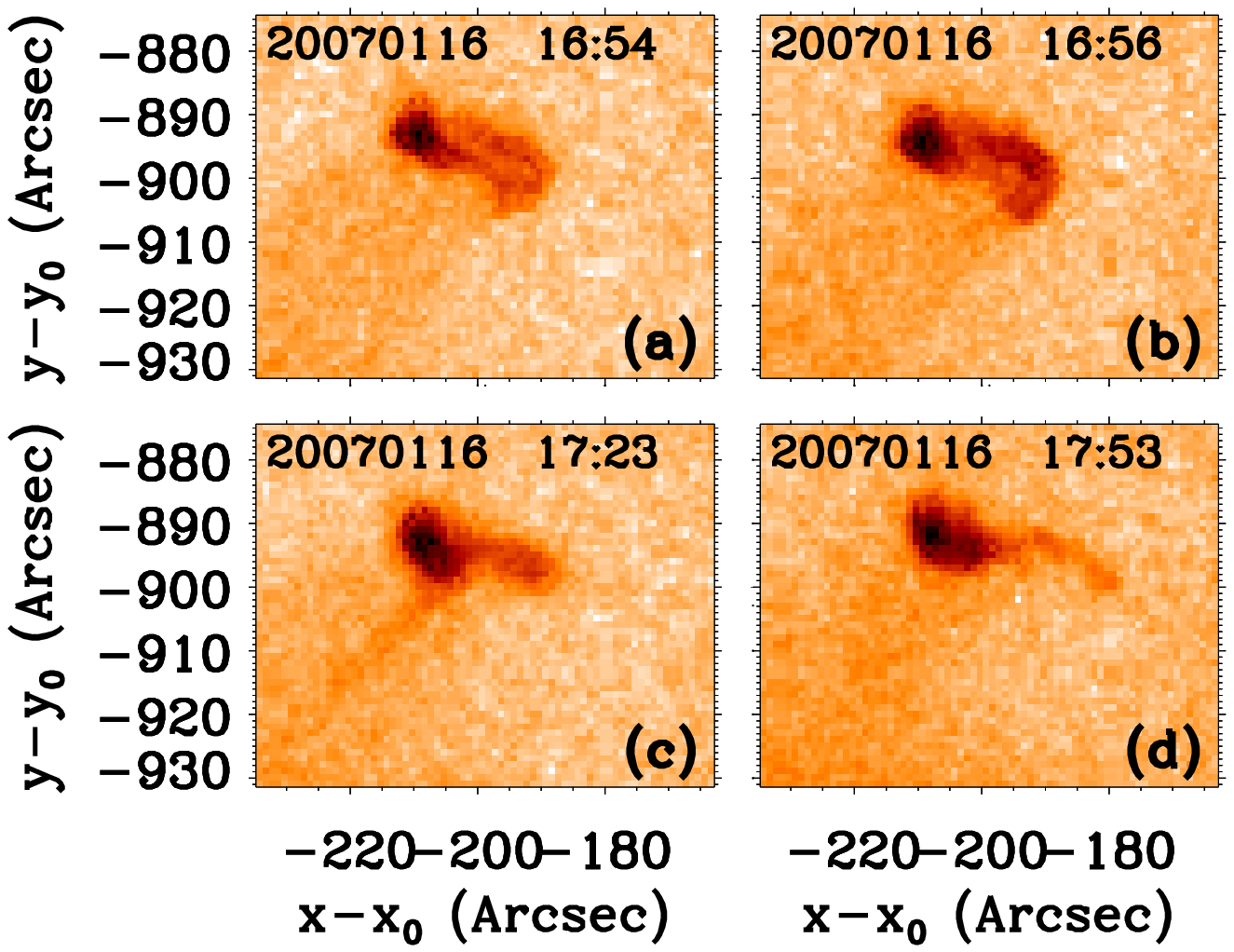}
\caption{Negative images showing the evolution of J-shaped loop systems (panel a) into a small-scale sigmoid that yielded multiple jets (see panels b and c). Notice also the shape change of the micro-sigmoid. The observations are recoreded with the XRT Al\_poly filter. \label{h3fig2}}
\end{center}
\end{figure}

Fig.~\ref{h3fig1} displays the different stages of evolution of an XBP located in the southern polar coronal hole on January 13, 2007. The initial appearance of the XBP resembles a system of dipolar loops (Fig.~\ref{h3fig1}a). The brightness of the XBP increased with time and its shape changed significantly. Fig.~\ref{h3fig1}b shows a section of the XBP extending westward (to the right). The newly appearing section is dim but clearly noticeable. The detailed structure of the XBP at its peak brightness (Fig.~\ref{h3fig1}c) is more complex and significantly different from its initial appearance. Fig.~\ref{h3fig1}c displays an S-shaped structure suggesting a coronal micro-sigmoid. The size of the micro-sigmoid is about 30{\arcsec} by 10{\arcsec}, which is approximately 10\% the size the large-scale sigmoid studied by McKenzie \& Canfield (2008). Like most large-scale, active-region and quiet-sun, sigmoids, this micro-sigmoid is not uniformly bright.

A jet erupted first on the western end of the sigmoid (Fig.~\ref{h3fig1}c). The jet subsequently appeared at slightly different locations along the sigmoidal structure (not shown here). In the aftermath of the jet eruption, the micro-sigmoid section where the jet erupted was altered and the brighter section became more diffuse (Fig.~\ref{h3fig1}d). In the course of the jet, the XBP was progressively losing its sigmoidal shape.

Fig.~\ref{h3fig2} shows another example of a polar micro-sigmoid leading to an X-ray jet. The XBP shown in Fig.~\ref{h3fig2}a is seemingly composed of J-shaped loops. These loops merged later to form one brighter sigmoid where a few asynchronous jets erupted from different locations along the sigmoid. This evolution is similar to the large scale case studied by Mckenzie \& Canfield (2008). Notice also the substantial morphological change of the source regions prior to and after the jet eruptions (Fig.~\ref{h3fig2}d).

\vspace{-0.5cm}
\section{Discussion}

High-resolution images recorded by XRT provide evidence for coronal X-ray jets emanating from small-scale (micro-) sigmoids at the polar coronal holes. The new data allowed us to follow in detail the evolution of the regions prior to X-ray jet eruptions. Our findings are important to understanding the nature of the physical processes that drive coronal jets and related features, such as polar plumes, H$\alpha$ surges, and spicules. Several characteristics of jets (e.g., untwisting and shapes) can now be explained.

Inadequate resolution led to the assumption that XBP structure is like a magnetic dipole and that the anemone model of jets provides an adequate description of XBP activity. It has been assumed that the emerging dipolar magnetic field reconnects with the open coronal background field and that the previously trapped plasma is expelled into a collimated beam presumably in open magnetic field lines. Although the anemone model has been shown to reproduce many morphological features of coronal jets, it also exhibits shortcomings in explaining other properties, such as helical structures of some of the jets.

The present observations of jets emanating from polar micro-sigmoids provide a natural explanation of the observed unwinding behavior of these structures. Eruptive coronal sigmoids have helical magnetic fields, regardless of their origin (i.e., flux rope or a sheared arcade). Following the eruption, helicity must be transported and redistributed over the entire post-eruption flux rope resulting in the unwinding motion of the erupting structure. $\lambda$-shaped jets may correspond to partially erupting micro-sigmoids. This is particularly clear if the jet occur on the dim end of the micro-sigmoid such as the case shown by Fig.~\ref{h3fig1}. Inverse-Y jets, on the other hand, may correspond to complete eruption of micro-sigmoids. For more details see Raouafi et al. (2010).

The findings reported here may lead to a more complete understanding of solar eruptive activity. The occurrence of small-scale sigmoidal structures in the solar atmosphere and their eruption - in the course of micro-flares - into X-ray jets may suggest that solar eruptive activity is self-similar over a wide range of scales. Self-similarity is not alien to the evolving Sun, even in cases of instabilities such as flares, and it is likely caused by an intrinsic self-organization that is due to the turbulent evolution in the solar atmosphere (e.g., Vlahos et al. 2002; Vlahos \& Georgoulis 2004; Georgoulis 2005).

This work is a first step toward uncovering a possible self-similarity in sigmoid sizes leading to a self-similarity in the respective CMEs, from micro-CMEs to the well-observed global CME events. It will be very interesting to determine whether the solar wind or coronal heating, for instance, are driven by self-similar physical processes acting on a wide spectrum of spatial scales.

%
%
%
%
%
%
%
%
%
%
%

\acknowledgments
Hinode is a Japanese mission developed and launched by ISAS/JAXA, with NAOJ as a domestic partner and NASA
and STFC (UK) as international partners. It is operated by these agencies in cooperation with the ESA and NSC
(Norway). N.-E.R's work is supported by NASA grant NNX08AJ10G. The work of the JHUALP authors in partially
supported by NASA grant number NNX08AJ10G.

\end{document}